\documentclass[showpacs,amssymb,amsmath,aps,prd,secnumarabic,twocolumn,groupedaddress]{revtex4-1}

\usepackage{graphicx}
\usepackage{comment}

\newcommand{\pt}{$p_{\rm T}$}
\newcommand{\kt}{$k_{\rm T}$}

\begin{document}

\title{ Model-independent constraints on the shape parameters \\ of dilepton angular
distributions }

\author{Pietro Faccioli$^a$, Carlos Louren\c{c}o$^b$, Jo\~ao
  Seixas$^{a,c}$, and Hermine K. W\"{o}hri$^{a,b}$}

\affiliation{$^a$Laborat\'orio de Instrumenta\c{c}\~ao e F\'{\i}sica Experimental de
  Part\'{\i}culas (LIP), 1000-149 Lisbon, Portugal\\
$^b$European Organization for Nuclear Research (CERN), 1211 Geneva 23, Switzerland\\
$^c$Physics Department, Instituto Superior T\'ecnico (IST), 1049-001 Lisbon, Portugal}

\date{\today}

\begin{abstract}

The coefficients determining the dilepton decay angular distribution of vector
particles obey certain positivity constraints and a rotation-invariant
identity. These relations are a direct consequence of the covariance properties
of angular momentum eigenstates and are independent of the production
mechanism.
The Lam--Tung relation can be derived as a particular case, simply
recognizing that the Drell--Yan dilepton is always produced transversely
polarized with respect to one or more quantization axes.
The dilepton angular distribution continues to be characterized by a
frame-independent identity also when the Lam--Tung relation is violated.
Moreover, the violation can be easily characterized by measuring a
one-dimensional distribution depending on one shape coefficient.

\end{abstract}



\maketitle

\sloppy

\section{Introduction}

Dilepton decay angular distributions directly reflect the average angular
momentum composition of the decaying state. Their measurements place strong
constraints on the characteristics and topology of the participating production
processes and can thus provide key information for the understanding of the
mechanisms of fundamental interactions.
In this paper we show how rotation covariance implies the existence of
completely general constraints on the coefficients of the dilepton decay
angular distribution of a $J=1$ particle. These constraints are valid for any
superposition of production mechanisms and are independent of the chosen
polarization frame. In particular, as first noted in Ref.~\cite{bib:LTGen},
the parameters characterizing the polar and
azimuthal anisotropies of the distribution satisfy a frame-independent
identity, directly reflecting a basic rotational property of $J=1$ angular
momentum eigenstates.
The well-known Lam--Tung relation~\cite{bib:LamTung}, a result specific to
Drell--Yan production in perturbative QCD, can be derived as a particular case
of this identity by simply noting that all subprocesses, up to $O(\alpha_s)$
contributions, produce transversely polarized dileptons, albeit with respect to
different quantization axes.
This result allows us to discern what in this relation embodies the dynamical
content of the specific processes involved and what reflects completely general
kinematic properties.
The existence of a frame-independent identity can be seen as a generalization
of the Lam--Tung relation. In fact, it is always possible to define a
frame-independent polarization observable, even when the Lam--Tung relation is violated
(or for processes different from Drell--Yan production). We also show that the
value of this observable (and, hence, possible violations of the Lam--Tung
relation) can be measured by simply determining a single-variable angular
distribution.
As an illustration of how simple and powerful the application of the
frame-independent formalism can be, we consider the significant violations of
the Lam--Tung identity measured in pion-nucleus experiments. The
intensively-studied possibility that these effects are caused by higher-order
corrections in perturbative-QCD is generally agreed to have been ruled out by
detailed calculations~\cite{bib:LToalpha2,bib:BergerQiu}. The same conclusion
can be reached in a much simpler way by considering rotational invariance and
symmetry properties.

\section{Angular distribution of dilepton decays of vector states}
\label{sec:ang}

\begin{figure}[htb]
\centering
\includegraphics[width=0.65\linewidth]{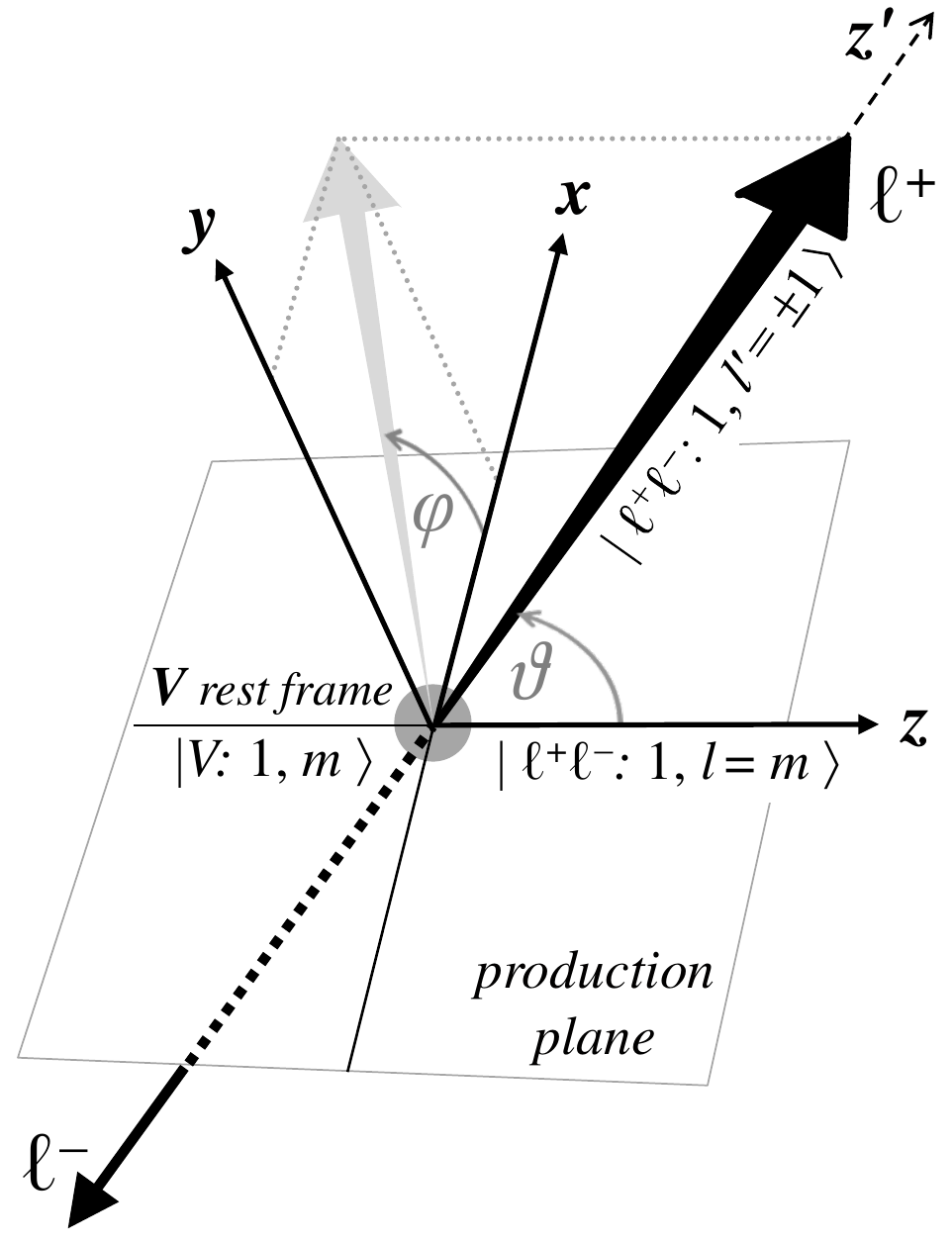}
\caption{\label{fig:notations} Sketch of the decay $V
  \rightarrow \ell^+ \ell^-$, showing the notations we use
  for axes, angles and angular momentum states. The $y$ and $z^\prime$ axes are oriented
towards the reader.}
\end{figure}

We start by expressing the observable dilepton angular distribution in a form
that keeps track of the angular momentum composition of the decaying state. We
study first the case of a single production ``subprocess'', here defined as a
process where the considered vector state $V$ is formed as a given
superposition of the three $J = 1$ eigenstates, $J_z = +1, -1, 0$ with respect
to a chosen polarization axis~$z$:
\begin{equation}
  | V \rangle =  b_{+1} \, |\hspace{-.2em}+\hspace{-.2em}1\rangle
+ b_{-1} \, |\hspace{-.2em}-\hspace{-.2em}1\rangle + b_{0} \, |0\rangle \,
  . \label{eq:state}
\end{equation}
The calculations are performed in the $V$ rest frame, where the common
direction of the two leptons define the reference axis $z^\prime$, oriented
conventionally along the direction of the positive lepton. The adopted
notations for axes, angles and angular momentum states are illustrated in
Fig.~\ref{fig:notations}.
%
%
We assume helicity conservation at the dilepton vertex, in the limit of
vanishing lepton masses. The dilepton system has thus angular momentum
projection $\pm 1$ along $z^\prime$, i.e.\ it is an eigenstate of
$J_{z^\prime}$, $|\ell^+ \ell^-; 1, l^\prime \rangle$, with $l^\prime = + 1$ or
$-1$.  This state can also be expressed as a superposition of eigenstates of
$J_z$, $|\ell^+ \ell^-; 1, l \rangle$ with $l = 0, \pm 1$, as
\begin{equation}
|\ell^+ \ell^-; 1, l^\prime \rangle = \sum_{l = 0, \pm 1} \mathcal{D}_{l
l^\prime}^{1}(\vartheta, \varphi) \, |\ell^+ \ell^-; 1, l \rangle \, ,
\label{eq:dilepton_state}
\end{equation}
where $\mathcal{D}_{l l^\prime}^{1}$ are complex coefficients describing the
rotation of a $J=1$ state from the set of axes $(x,y,z)$ to the set
$(x^\prime,y^\prime,z^\prime)$~\cite{bib:BrinkSatchler},
\begin{equation}
\mathcal{D}_{l l^\prime}^{1}(\vartheta, \varphi) = e^{i (l^\prime-l) \varphi }
d_{l l^\prime}^{1}(\vartheta) \, , \label{eq:D_matrix}
\end{equation}
with
\begin{equation}
\begin{split}
d^1_{0, \pm 1} = \pm & \sin \vartheta / \sqrt{2} \, , \;\; d^1_{\pm 1, \pm 1} =
(1 + \cos \vartheta) / 2 \, , \\
& d^1_{\pm 1, \mp 1} = (1 - \cos \vartheta) / 2 \, .
\label{eq:reduced_d_matrix}
\end{split}
\end{equation}

The amplitude of the partial process $V(m) \rightarrow  \ell^+ \ell^-
(l^\prime)$ represented in Fig.~\ref{fig:notations} is
\begin{eqnarray}
B_{m l^\prime} \; & = & \; \sum_{l = 0, \pm 1} \mathcal{D}_{l l^\prime}^{1
*}(\vartheta, \varphi) \, \langle \ell^+ \ell^-; 1, l \; | \,
\mathcal{B} \, | \;
V; 1, m \rangle \nonumber \\[2mm]
& = & \; B \; \mathcal{D}_{m l^\prime}^{1 *}(\vartheta, \varphi) \, ,
\label{eq:jpsi_to_ll_amplitude}
\end{eqnarray}
where we imposed that the transition operator $\mathcal{B}$ is of the form
$\langle \ell^+ \ell^-; 1, l \; | \, \mathcal{B} \, | \; V; 1, m \rangle = B \,
\delta_{m \, l}$ because of angular momentum conservation, with $B$ independent
of $m$ (for rotational invariance). The total amplitude for $V \rightarrow
\ell^+ \ell^- (l^\prime)$, where $V$ is given by the superposition written in
Eq.~\ref{eq:state}, is
\begin{eqnarray}
B_{l^\prime} \; & = & \; \sum_{m = 0, \pm 1} b_m B \; \mathcal{D}_{m
l^\prime}^{1
*}(\vartheta, \varphi) \nonumber \\[2mm]
& = & \; \sum_{m = 0, \pm 1} a_m \; \mathcal{D}_{m l^\prime}^{1
*}(\vartheta, \varphi) \, .
\label{eq:jpsi_to_ll_fullamplitude}
\end{eqnarray}
The probability of the transition is obtained by squaring
Eq.~\ref{eq:jpsi_to_ll_fullamplitude} and summing over the (unobserved) spin
alignments ($l^\prime = \pm 1$) of the dilepton system, with equal weights
attributed, for parity conservation, to the two configurations.  Using
Eqs.~\ref{eq:D_matrix} and~\ref{eq:reduced_d_matrix} one finally obtains the
angular distribution
\begin{eqnarray}
W(\cos \vartheta, \varphi)  & \propto &  \sum_{l^\prime = \pm 1} | B_{l^\prime} |^2 
                             \; \propto \; \frac{\mathcal{N}}{(3 +
\lambda_{\vartheta})}\; (1 + \lambda_{\vartheta} \cos^2\!\vartheta \nonumber \\
& + & \;\lambda_{\varphi} \sin^2 \vartheta \cos 2 \varphi\; +\;
\lambda_{\vartheta \varphi} \sin 2 \vartheta \cos \varphi
\label{eq:ang_distr_subproc} \\
& + & \;\lambda^{\bot}_{\varphi} \sin^2 \vartheta \sin 2 \varphi \;+\;
\lambda^{\bot}_{\vartheta \varphi} \sin 2 \vartheta \sin \varphi  ) \, ,
\nonumber
\end{eqnarray}
with $\mathcal{N} = |a_0|^2 + |a_{+1}|^2 + |a_{-1}|^2$ and
\begin{eqnarray}
  \lambda_{\vartheta} & = & \frac{{\mathcal{N}}-3 |a_0|^2}{\mathcal{N}+|a_0|^2}  \, ,
  \nonumber \\
  \lambda_{\varphi}  & = & \frac{ 2 \, \mathrm{Re} [a_{+1}^{*}
    a_{-1}] }{\mathcal{N}+|a_0|^2} \, , \nonumber \\
  \lambda_{\vartheta \varphi} &  = & \frac{ \sqrt{2} \, \mathrm{Re} [ a_{0}^{*} ( a_{+1} - a_{-1})] }{\mathcal{N}+|a_0|^2} \,
  , \label{eq:lambdas_vs_amplitudes} \\
  \lambda^{\bot}_{\varphi}  & = & \frac{ 2 \, \mathrm{Im} [a_{+1}^{*} a_{-1}] }{\mathcal{N}+|a_0|^2} \, , \nonumber \\
  \lambda^{\bot}_{\vartheta \varphi} &  = & \frac{ - \sqrt{2} \,
    \mathrm{Im} [a_{0}^{*} (a_{+1} + a_{-1})] }{\mathcal{N}+|a_0|^2} \, . \nonumber
\end{eqnarray}

In this paper we consider inclusive production. Therefore, for all of the
popular choices of frame, the $xz$ plane coincides with the production plane,
containing the directions of the colliding particles and of the decaying
particle itself. The last two terms in Eq.~\ref{eq:ang_distr_subproc} introduce
an asymmetry of the distribution by reflection with respect to the production
plane, an asymmetry which is not forbidden in individual events. In hadronic
collisions, due to the intrinsic parton transverse momenta, for example, the
``natural'' polarization plane does \emph{not} coincide event-by-event with the
experimental production plane.
However, the symmetry by reflection must be a property of the observed
\emph{event distribution}, integrating over many events,
when only parity-conserving processes contribute.
Indeed, the terms in $\sin^2 \vartheta \sin 2 \varphi$ and $\sin 2 \vartheta
\sin \varphi$ are unobservable, because they vanish on average.
In the presence of $n$ contributing production processes with weights
$f^{(i)}$, the most general \emph{observable} distribution can be written as
\begin{eqnarray}
  W(\cos \vartheta, \varphi) \, & = & \, \sum_{i = 1}^{n} f^{(i)}
  W^{(i)}(\cos \vartheta, \varphi) \nonumber \\
  & \propto & \, \frac{1}{(3 + \lambda_{\vartheta})} \,
  (1 + \lambda_{\vartheta} \cos^2\!\vartheta \label{eq:observable_ang_distr} \\
  & + &  \lambda_{\varphi} \sin^2 \vartheta \cos 2 \varphi +
  \lambda_{\vartheta \varphi} \sin 2 \vartheta \cos \varphi ) \, ,\nonumber
\end{eqnarray}
where $W^{(i)}(\cos \vartheta, \varphi)$ is the ``elementary'' decay
distribution corresponding to a single subprocess (given by
Eqs.~\ref{eq:ang_distr_subproc} and~\ref{eq:lambdas_vs_amplitudes}, adding the
index $(i)$ to the decay parameters). Each of the three observable shape
parameters, $X = \lambda_{\vartheta}$, $\lambda_{\varphi}$ and
$\lambda_{\vartheta \varphi}$, is a weighted average of the corresponding
parameters, $X^{(i)}$, characterizing the single subprocesses,
%
%
\begin{equation}
  X \, = \, \frac{ \sum_{i = 1}^{n} g^{(i)} \,  X^{(i)} }{ \sum_{i = 1}^{n} g^{(i)} } \, ,
\label{eq:parameters}
\end{equation}
with $g^{(i)} =   f^{(i)} \mathcal{N}^{(i)} / (3 + \lambda_{\vartheta}^{(i)})$.

\section{Positivity constraints}
\label{sec:constraints}

\begin{figure}[ht]
\centering
\includegraphics[width=1.0\linewidth]{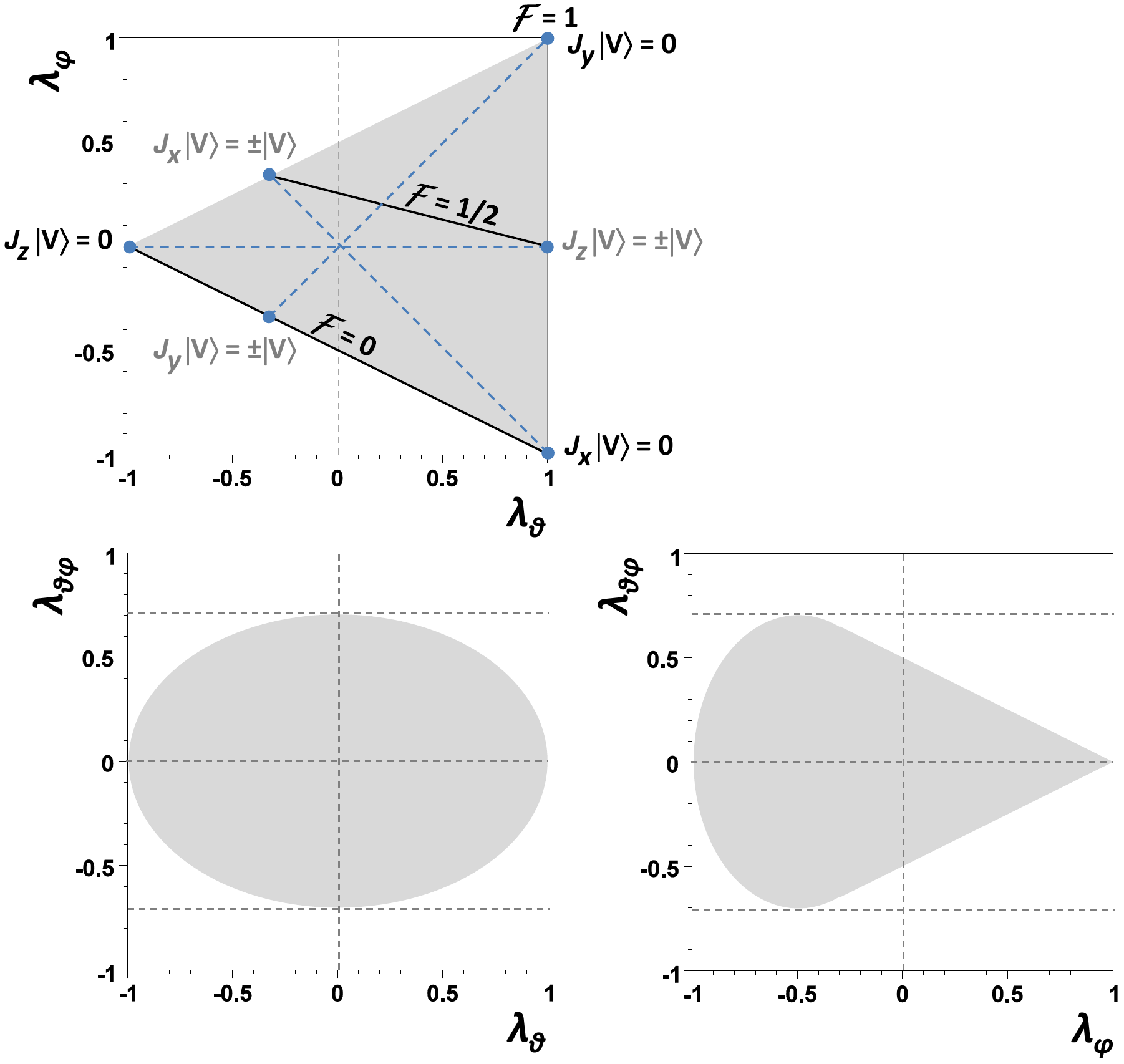}
\caption{\label{fig:triangles} Allowed regions for the decay angular parameters
(shaded areas). The upper plot also indicates the points corresponding to pure
angular momentum configurations and to specific values of the
rotation-invariant observable $\mathcal{F}$, introduced in
Section~\ref{sec:invariant}. }
\end{figure}

Equation~\ref{eq:lambdas_vs_amplitudes} implies the relations
\begin{eqnarray}
1 \pm \lambda_{\varphi}^{(i)} & = & ( |a_{+1}^{(i)} \pm a_{-1}^{(i)}|^2 + 2
|a_{0}^{(i)}|^2) / ( \mathcal{N}^{(i)} + |a_{0}^{(i)}|^2 ) \, ,\nonumber \\
\lambda_{\vartheta}^{(i)} \pm \lambda_{\varphi}^{(i)} & = & ( |a_{+1}^{(i)} \pm
a_{-1}^{(i)}|^2 - 2 |a_{0}^{(i)}|^2) / ( \mathcal{N}^{(i)} + |a_{0}^{(i)}|^2 ) \, ,\nonumber \\
|\lambda_{\vartheta \varphi}^{(i)}| & \le & \sqrt{2} |a_{0}^{(i)}| |a_{+1}^{(i)} - a_{-1}^{(i)}| / ( \mathcal{N}^{(i)} + |a_{0}^{(i)}|^2 ) \, ,\nonumber \\
|\lambda_{\vartheta \varphi}^{\bot (i)}| & \le & \sqrt{2} |a_{0}^{(i)}|
|a_{+1}^{(i)} + a_{-1}^{(i)}| / ( \mathcal{N}^{(i)} + |a_{0}^{(i)}|^2 ) \, ,
\label{eq:relations_subprocess}
\end{eqnarray}
where the index $(i)$ now explicitly denotes the single-subprocess quantities.
Equation~\ref{eq:relations_subprocess} implies the following
relations between the coefficients of the angular distribution:
\begin{eqnarray}
(1 - \lambda_{\varphi}^{(i)})^2 - (\lambda_{\vartheta}^{(i)} -
\lambda_{\varphi}^{(i)})^2 & \ge & 4 \lambda_{\vartheta \varphi}^{(i)2} \, ,\nonumber \\
(1 ´+ \lambda_{\varphi}^{(i)})^2 - (\lambda_{\vartheta}^{(i)} +
\lambda_{\varphi}^{(i)})^2 & \ge & 4 \lambda_{\vartheta \varphi}^{\bot (i)2} \, .
\label{eq:triangle_inequalities}
\end{eqnarray}
From these expressions we finally reach the following set of inequalities:
\begin{eqnarray}
&&|\lambda_\varphi| \le \frac{1}{2}\, (1 + \lambda_\vartheta ) \, , \quad
\lambda_\vartheta^2 + 2 \lambda_{\vartheta \varphi}^2 \le 1 \, , \nonumber \\
&& |\lambda_{\vartheta \varphi}| \le \frac{1}{2}\, (1 - \lambda_\varphi ) \, , \label{eq:triangles}\\
&& (1 + 2 \lambda_\varphi)^2 + 2\lambda_{\vartheta \varphi}^2 \le 1 \;\;\;
\mathrm{for} \;\;\; \lambda_\varphi < -1/3 \,  . \nonumber
\end{eqnarray}
Here we have dropped the index $(i)$ because these relations are completely
general and valid for any superposition of production processes, as can be
verified using Eq.~\ref{eq:parameters} (being $g^{(i)} > 0$) and, for the two
quadratic relations, the Schwarz inequality,
\begin{equation}
  \left( \frac{ \sum_{i = 1}^{n} g^{(i)} \,
  X^{(i)} }{ \sum_{i = 1}^{n} g^{(i)} } \right)^2 \; \le \;
  \frac{ \sum_{i = 1}^{n} g^{(i)} \,
  X^{(i)2} }{ \sum_{i = 1}^{n} g^{(i)} } \, .
\label{eq:schwarz}
\end{equation}
Equation~\ref{eq:triangles} implies, for example, $|\lambda_\varphi| \le 1 $,
$|\lambda_{\vartheta \varphi}| \le \sqrt{2}/2$, $|\lambda_\varphi| \le 0.5$ for
$\lambda_{\vartheta} = 0$ and $\lambda_\varphi \to 0$ for $\lambda_{\vartheta} \to
-1$.
There is an alternative notation, widespread in the literature, where the
coefficients $\lambda$, $\nu/2$ and $\mu$ replace, respectively,
$\lambda_{\vartheta}$, $\lambda_{\varphi}$ and $\lambda_{\vartheta
  \varphi}$. In that case, hence, we have $|\nu| \le 2$.
The most general domain for the three angular parameters is represented in
Fig.~\ref{fig:triangles}.
%
%
The upper plot also illustrates the meaning of specific points of the
$\lambda_{\vartheta},\lambda_{\varphi}$ plane in terms of angular momentum
state of the decaying particle. The six points indicated on the border of the
triangle are the combinations of observable parameters corresponding to pure
eigenstates of $J_x$, $J_y$ and $J_z$ with eigenvalues 0 or $\pm 1$.
In particular, the three vertices represent \emph{univocally} the
well-defined cases in which \emph{all} contributing production processes lead
to the same, fully \emph{longitudinal} polarization along the
$x$, $y$ or $z$ axes. The three points lying on the sides
of the triangle, however, can either be the result of purely \emph{transverse} polarizations
along the $x$, $y$ and $z$ axes, or of suitable mixtures of angular
momentum eigenstates and/or superpositions of different processes, polarized along
different axes.

\section{Polarization-frame-independent observable}
\label{sec:invariant}

The rotation-covariance properties of the generic $J=1$ state defined
in Eq.~\ref{eq:state} imply two propositions.

\medskip
\begingroup
\parindent=0pt
$\bullet$ Proposition 1: The amplitude combination $b_{+1}+b_{-1}$ is invariant
by rotation around the $y$ axis.
\smallskip

$\bullet$ Proposition 2: There exists a quantization axis $z^{\star}$ with
respect to which $b_0^{\star}=0$; if $b_0$, $b_{+1}$ and $b_{-1}$ are real,
$z^{\star}$ belongs to the $xz$ plane.
\medskip
\endgroup

In fact, for successive rotations about, respectively, the $z$ and $y$ axes by
angles $\varphi$ and $\vartheta$, a pure $J=1, J_z$ angular momentum eigenstate
$|m \rangle$ transforms according to the relation (analogous to
Eq.~\ref{eq:dilepton_state}, but describing the inverse rotation)
\begin{equation}
|m \rangle = \sum_{m^\prime = 0, \pm 1} \mathcal{D}_{m m^\prime}^{1
*}(\vartheta, \varphi) \, |m^\prime \rangle \, .
\label{eq:eigenstate_rotation}
\end{equation}
In the basis of the rotated eigenspace, the state in Eq.~\ref{eq:state} has
components
\begin{equation}
b_k^\prime  = \sum_{m = 0, \pm 1} b_m \mathcal{D}_{m k}^{1*}(\vartheta,
\varphi) \, . \label{eq:V_rotated_components}
\end{equation}
For a rotation in the production plane (about $y$: $\varphi = 0$),
\begin{eqnarray}
b_{+1}^\prime + b_{-1}^\prime & = & \sum_{m = 0, \pm 1} b_m [d_{m
,+1}^{1}(\vartheta) + d_{m ,-1}^{1}(\vartheta)] \nonumber \\
& = & b_{+1} + b_{-1} \, , \label{eq:proof_of_proposition1}
\end{eqnarray}
where we have used (Eq.~\ref{eq:reduced_d_matrix}) $d_{\pm 1,
+1}^{1}(\vartheta) + d_{\pm 1,-1}^{1}(\vartheta) = 1$, $d_{0,
+1}^{1}(\vartheta) + d_{0,-1}^{1}(\vartheta) = 0$. This proves Proposition~1.
We now address Proposition~2 taking $|V^{(i)} \rangle$ defined with real $b_0$
(always possible). After a generic rotation, the zero-helicity component
becomes (Eqs.~\ref{eq:V_rotated_components}, \ref{eq:D_matrix},
\ref{eq:reduced_d_matrix})
\begin{equation}
b_{0}^\prime(\vartheta,\varphi) \; = \; b_0 \cos \vartheta - \frac{1}{\sqrt{2}}
(b_{+1} e^{i \varphi} - b_{-1} e^{-i \varphi}) \sin \vartheta \, .
\label{eq:proof_of_proposition2}
\end{equation}
It can be verified explicitly that the equation
$b_{0}^\prime(\vartheta,\varphi) = 0$ has always a solution, given by
\begin{align}
\begin{split}
  \cos \vartheta^{\star} & = \frac{R_+ R_- + I_+ I_-}{\sqrt{2 b_0^{2}
      (R_+^2 + I_-^2) +
      (R_+ R_- + I_+ I_-)^2 } } \, , \\
  \cos \varphi^{\star} & = \frac{R_+}{\sqrt{R_+^2 + I_-^2}} \, , \quad
  \sin \varphi^{\star} = - \frac{I_-}{\sqrt{R_+^2 + I_-^2}} \, ,
\end{split}
\end{align}
where $R_{\pm} = \mathrm{Re}(b_{+1} \pm b_{-1})$ and $I_{\pm} =
\mathrm{Im}(b_{+1} \pm b_{-1})$. If all three amplitudes are real, then
$\varphi^{\star} = 0$ and the rotation is around the $y$ axis.

We remind that the decay amplitudes $a_m$ are simply proportional to the
angular momentum components $b_m$. Therefore, Proposition~1 and the obvious
rotation invariance of $|a_0|^2 +|a_{+1}|^2 + |a_{-1}|^2$ imply that, for each
subprocess $(i)$, the quantity
\begin{equation}
  \mathcal{F}^{(i)}=\frac{1}{2} \,
  \frac{|a_{+1}^{(i)}+a_{-1}^{(i)}|^2}{|a_0^{(i)}|^2
    + |a_{+1}^{(i)}|^2 + |a_{-1}^{(i)}|^2}
  \label{eq:F_i}
\end{equation}
(included between 0 and 1) is independent of the chosen frame. Using also
Eqs.~\ref{eq:lambdas_vs_amplitudes} and~\ref{eq:parameters}, we find that the
following combination of observable parameters of the dilepton decay
distribution is frame-independent (invariant by rotation about the $y$ axis):
\begin{equation}
  \mathcal{F} \, = \, \frac{ \sum_{i = 1}^{n} f^{(i)}
    \mathcal{N}^{(i)} \mathcal{F}^{(i)}}{\sum_{i =
      1}^{n} f^{(i)} \mathcal{N}^{(i)}}  \, = \, \frac{1 +
    \lambda_\vartheta + 2
    \lambda_\varphi}{3 + \lambda_\vartheta} \, . \label{eq:F}
\end{equation}
The upper plot in Fig.~\ref{fig:triangles} shows the loci of points in the
$\lambda_\vartheta,\lambda_\varphi$ plane corresponding to $\mathcal{F} = 0$,
$\mathcal{F} = 1/2$ and $\mathcal{F} = 1$. The $\mathcal{F} = 0$ and
$\mathcal{F} = 1/2$ lines include the cases of, respectively, full longitudinal
and full transverse polarizations with respect to any axis belonging to the
production plane. The uniquely defined $\mathcal{F} = 1$ point corresponds to
the theoretical case of a full longitudinal polarization along the $y$ axis.

We mention, for completeness, that the quantity
\begin{equation}
  \mathcal{G} \, = \, \frac{ \sum_{i = 1}^{n} f^{(i)}
    \mathcal{N}^{(i)} \mathcal{G}^{(i)}}{\sum_{i =
      1}^{n} f^{(i)} \mathcal{N}^{(i)}}  \, = \, \frac{1 +
    \lambda_\vartheta - 2
    \lambda_\varphi}{3 + \lambda_\vartheta} \, , \label{eq:G}
\end{equation}
with
\begin{equation}
  \mathcal{G}^{(i)}=\frac{1}{2} \,
  \frac{|a_{+1}^{(i)}-a_{-1}^{(i)}|^2}{|a_0^{(i)}|^2
    + |a_{+1}^{(i)}|^2 + |a_{-1}^{(i)}|^2} \, ,
  \label{eq:G_i}
\end{equation}
is invariant by rotation about the $x$ axis. Finally, the parameter
$\lambda_\vartheta$ itself is invariant by rotation about $z$.

\section{Polarization-frame-independent angular distribution}
\label{sec:cosalpha}

Clearly, we can determine the frame-invariant polarization observable
$\mathcal{F}$ through the measurement of the two-dimensional, three-parameters
angular distribution of Eq.~\ref{eq:observable_ang_distr}. This procedure is
particularly useful when performed in two sufficiently different reference
frames, to probe systematic effects caused by experimental
biases~\cite{bib:ImprovedQQbarPol}, since different values of
$\lambda_\vartheta$ and $\lambda_\varphi$, but identical values of
$\mathcal{F}$, are expected in each frame. However, it may be convenient to
determine $\mathcal{F}$ directly from a one-dimensional, single-parameter
angular distribution. The distribution itself must be, like $\mathcal{F}$,
invariant by rotation about the $y$ axis. This restricts the possibilities for
the definition of the corresponding angular variable to
\begin{equation}
  \cos \alpha \; = \; \sin \vartheta \, \sin \varphi \, ,
  \label{eq:cosalpha}
\end{equation}
where $\alpha$ is the angle formed by the lepton with the $y$ axis. The $\cos
\alpha$ distribution must be of the form
\begin{equation}
  w(\cos\alpha) \; \propto \; 1 + \lambda_\alpha \cos^2\!\alpha \, ,
  \label{eq:cosalphadistr}
\end{equation}
as any parity-conserving distribution of the angle formed with respect to an
axis, when only $J=1$ wave functions are involved. The relation of
$\lambda_\alpha$ to $\lambda_\vartheta$, $\lambda_\varphi$ and $\mathcal{F}$
can be found by imposing the condition
\begin{eqnarray}
  \langle \cos^2\!\alpha \rangle & = & \int_{-1}^{+1} \! \cos^2\!\alpha \,\, w(\cos \alpha) \, \mathrm{d}(\cos \alpha) \label{eq:lambdaalphaderivation} \\
  & = &  \int_{0}^{2 \pi} \! \int_{-1}^{+1} \! (\sin \vartheta \, \sin \varphi)^2 \, W(\cos \vartheta, \varphi) \, \mathrm{d}(\cos \vartheta) \mathrm{d}\varphi  \,
  ,
  \nonumber
\end{eqnarray}
with the result
\begin{equation}
  \lambda_\alpha \; = \; - \frac{\lambda_\vartheta + 3 \lambda_\varphi}{2+ \lambda_\vartheta +
  \lambda_\varphi} \; = \; \frac{1 - 3 \mathcal{F}}{1 + \mathcal{F}}
  \, .
  \label{eq:lambdaalpha}
\end{equation}
%

\section{The Lam--Tung relation as a particular case}
\label{sec:LamTung}

It has been noticed long ago that, in the case of Drell--Yan production, the
shape parameters $\lambda_\vartheta$ and $\lambda_\varphi$ obey the
frame-independent expression $\lambda_\vartheta \, + \, 4 \, \lambda_\varphi \,
= \, 1$, commonly known as the ``Lam--Tung relation''~\cite{bib:LamTung}.
Although the dilepton production cross section is substantially modified by QCD
corrections, the \textit{relation} between the different helicity contributions
to this cross section remains unchanged up to $O(\alpha_s)$, a seemingly
surprising feature. Relatively small corrections affect the angular
distribution when subsequent orders in $\alpha_s$ are taken into
account~\cite{bib:LToalpha2,bib:BergerQiu}. Given its robustness within
perturbative QCD, deviations from the Lam--Tung relation have been considered as
a signal of higher twist contributions~\cite{bib:highertwist} or
non-perturbative effects caused by intrinsic parton \kt~\cite{bib:kt}, or even
parton saturation~\cite{bib:lowx}.

\begin{figure}[htb]
\centering
\includegraphics[width=\linewidth]{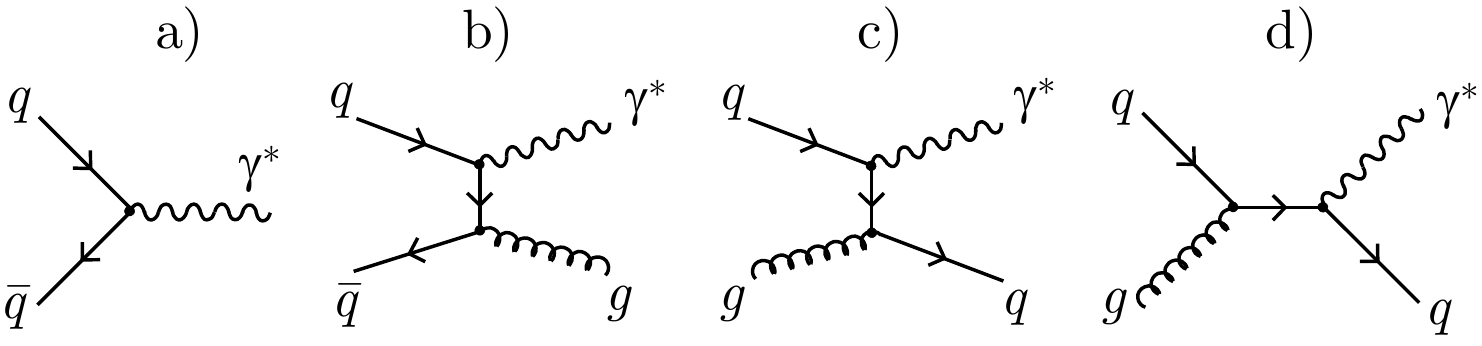}
\caption{$O(\alpha_s^0)$ and $O(\alpha_s^1)$ processes for Drell--Yan
  production, giving rise to transverse dilepton polarizations along
  different quantization axes: Collins--Soper (a), Gottfried--Jackson
  (b, c) and helicity (d).}
\label{fig:DYprocesses}
\end{figure}

Actually, the Lam--Tung relation is a particular case of
the more general invariant relation presented in Eq.~\ref{eq:F}.
Indeed, in Drell--Yan production up to $O(\alpha_s)$, neglecting parton
transverse momenta, the topology of each contributing subprocess
(quark-antiquark annihilation without or with single gluon emission,
Compton-like quark-gluon scattering, etc.) is characterized by one
reaction plane, coinciding with the experimental production plane.
Therefore, as mentioned in Section~\ref{sec:ang}, we can set $\lambda^{\bot
  (i)}_{\varphi} = \lambda^{\bot (i)}_{\vartheta \varphi} = 0$ for
each single subprocess, $(i)$. Imposing this condition in
Eq.~\ref{eq:lambdas_vs_amplitudes}, we find that the three partial decay
amplitudes, $a_m^{(i)}$, and, therefore, the corresponding angular momentum
components, $b_m^{(i)}$, have the same complex phase. Proposition~2 implies,
then, that the observed dilepton distribution is a superposition of
sub-distributions characterized by
\begin{equation}
\lambda^{(i)\star}_\vartheta = +1 \, , \quad \lambda^{(i)\star}_\varphi =
2\,\mathcal{F}^{(i)}-1 \, , \quad \lambda^{(i)\star}_{\vartheta \varphi} = 0 \, ,
\label{eq:elementaryDistr_LT}
\end{equation}
each one referred to a specific polarization axis $z^{(i)\star}$ belonging
to the production plane.
Assuming helicity conservation at the production vertex (i.e., that the
participating quarks are massless), the zero-order quark-antiquark annihilation
process in Drell--Yan production, Fig.~\ref{fig:DYprocesses}\,(a), leads to a
decay anisotropy of the kind $1+\cos^2\!\vartheta$ with respect to the
direction of the relative momentum between quark and antiquark, experimentally
approximated by the Collins--Soper (CS) frame~\cite{bib:coll_sop}.
In the $O(\alpha_s)$ processes, on the other hand, the photon couples
to one real quark and to the intermediate virtual quark, this latter
having a well-defined momentum.
Also in this case helicity conservation leads to a decay anisotropy of the kind
$1+ \cos^2\!\vartheta$, but now with respect to the direction of the relative
momentum between the real and virtual quarks.
Experimentally, this quantization axis corresponds to the
Gottfried--Jackson (GJ,~\cite{bib:gott_jack}) axis for the processes
presented in Fig.~\ref{fig:DYprocesses}\,(b)-(c), and to the helicity
axis for the process shown in Fig.~\ref{fig:DYprocesses}\,(d).

%
%
Effectively, therefore, all subprocesses contributing to Drell--Yan production
up to $O(\alpha_s)$ lead individually to the same kind of \emph{fully
transverse, purely polar} decay anisotropy, even if \emph{with respect to
three different ``natural'' axes, $z^{(i)\star}$}. In each case
$\lambda^{(i)\star}_\varphi = 0$, meaning that
$\mathcal{F}^{(i)} = 1/2$ for all subprocesses (Eq.~\ref{eq:elementaryDistr_LT})
and, thus, implying $\mathcal{F} = 1/2$.
This latter equation coincides with the Lam--Tung relation.
In other words, we have shown that the \emph{frame independence} of the
Lam--Tung relation is a simple \emph{kinematic} consequence of the rotational
properties of the $J=1$ angular momentum eigenstates (leading, in general, to
Eq.~\ref{eq:F}), while its specific \emph{form} ($\mathcal{F}=1/2$) derives
from the \emph{dynamical} input that all contributing subprocesses produce
\emph{transversely} polarized states.

Deviations from the Lam--Tung relation are often parametrized in terms of the
quantity $\Delta = \lambda_{\vartheta} \, + \, 4 \, \lambda_{\varphi} - 1$ in
the experimental and theoretical literature. However, the correspondingly
assumed relation $ \lambda_{\vartheta} \, + \, 4 \, \lambda_{\varphi} = 1 +
\Delta$ is not frame-invariant (it cannot be rewritten in the form of
Eq.~\ref{eq:F} for a certain value of $\mathcal{F}$) and, hence, the
``violation level'' expressed by $\Delta$ depends on the frames used in the
analyses. We propose that future searches for violations of the Lam--Tung
relation evaluate the (frame-invariant) deviation of $\mathcal{F}$ from $1/2$.

Following the considerations of Section~\ref{sec:cosalpha}, tests of the
Lam--Tung relation can be performed by simply determining the $\cos \alpha$
distribution and measuring the deviation of $\lambda_\alpha$ from the value
$-1/3$. Vice versa, in regimes where the validity of the Lam--Tung relation (or,
more generally, the intrinsic transverseness of the polarization) can be
considered as a characterizing feature of the physical process under study, the
event distribution
\begin{equation}
  w(\cos\alpha) \; \propto \; 1 - \frac{1}{3} \cos^2\!\alpha
  \label{eq:cosalphadistr_LamTung}
\end{equation}
can be used to check the purity of the selected signal sample and/or to provide an
event-by-event criterium for signal-background discrimination.

\section{Violation of the Lam--Tung relation in pion-nucleus data}
\label{sec:pions}

\begin{figure}[t]
\centering
\includegraphics[width=0.54\linewidth]{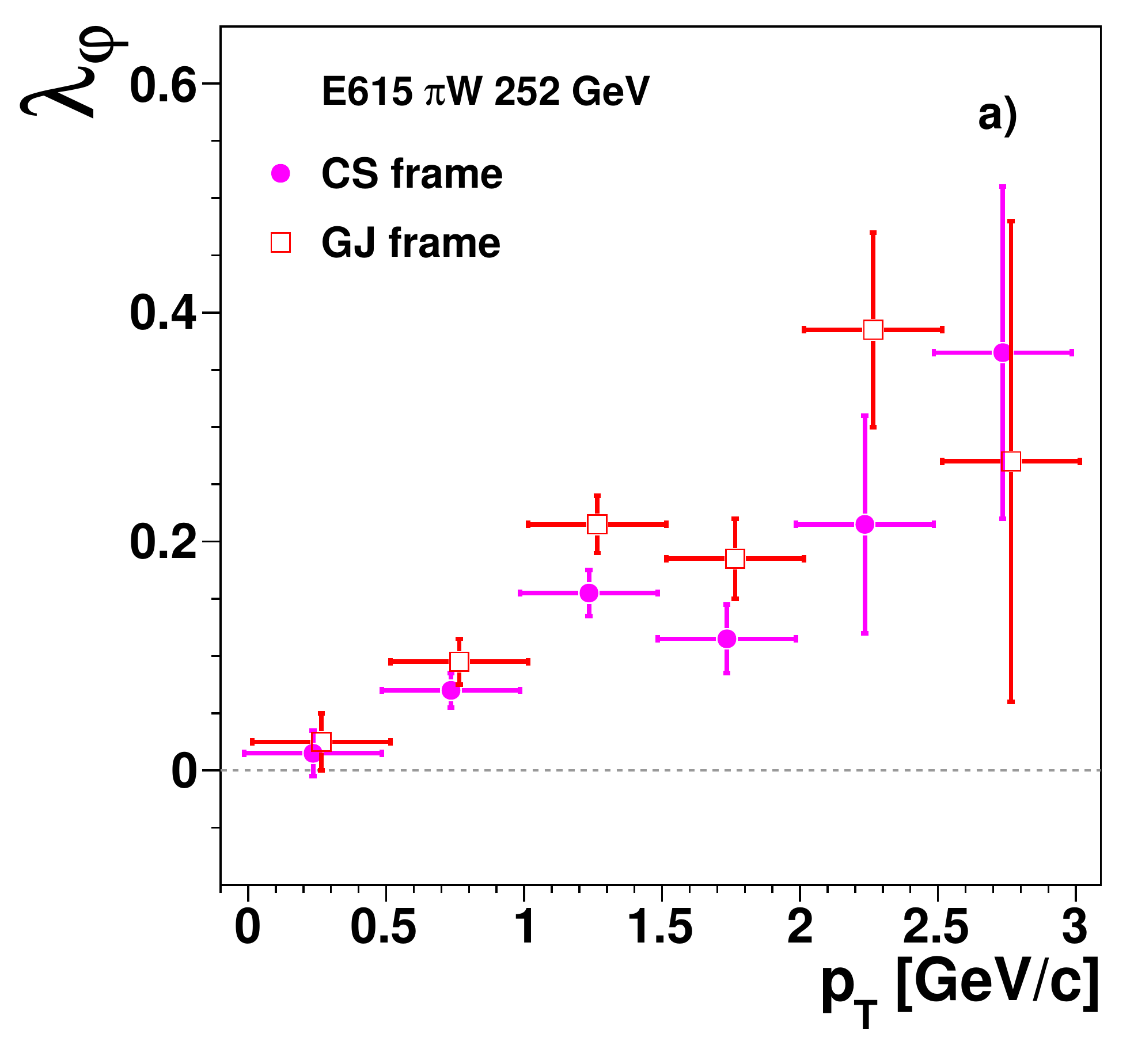}
\includegraphics[width=0.54\linewidth]{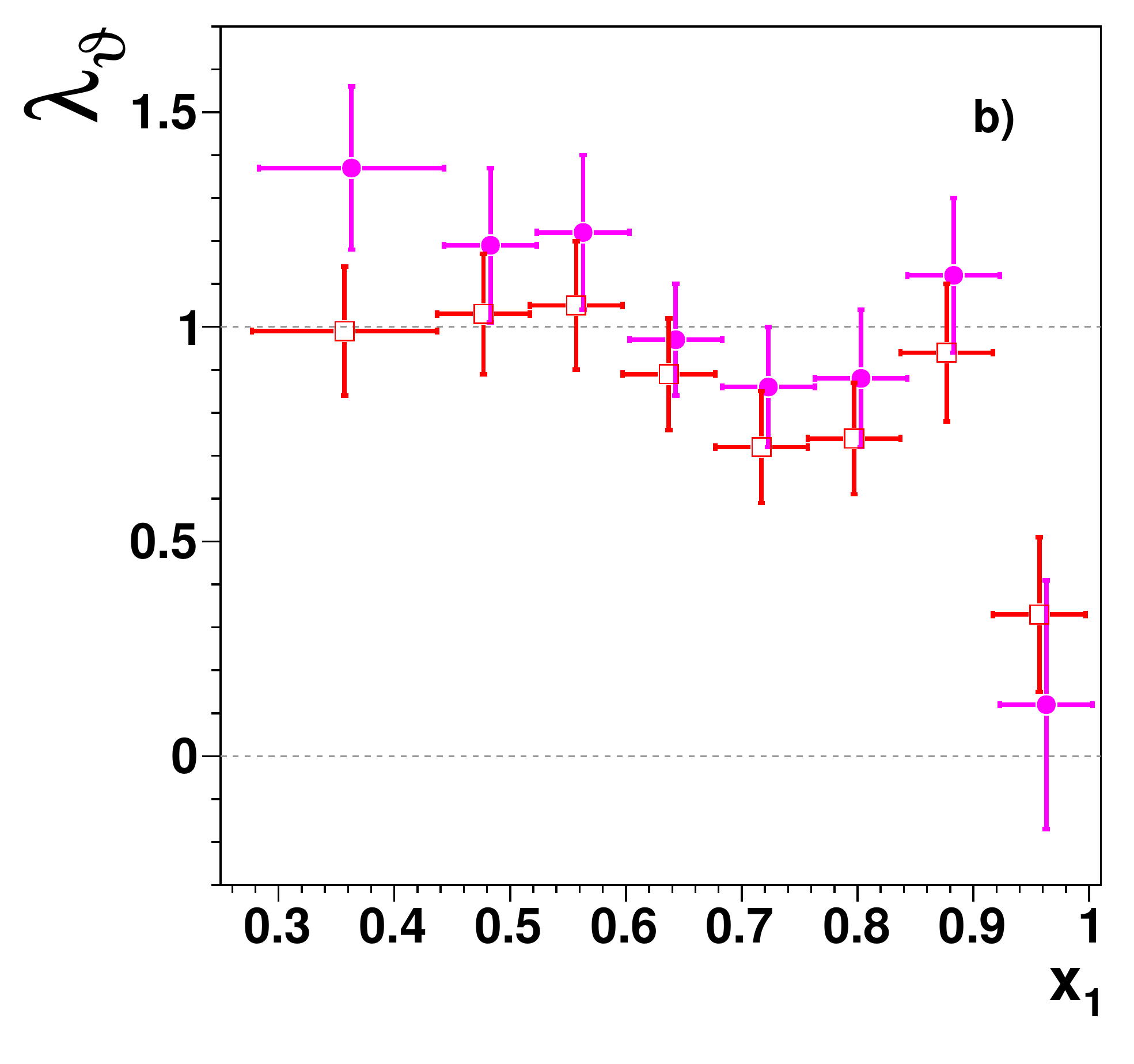}
\caption{The E615 measurements of the Drell--Yan azimuthal anisotropy as a
function of \pt\ (a) and of the polar anisotropy as a function of $x_1$ (b).
The points are slightly displaced in the horizontal axis for improved
visibility.} \label{fig:DYdata_lambdas}
\end{figure}

\begin{figure}[!ht]
\centering
\includegraphics[width=0.54\linewidth]{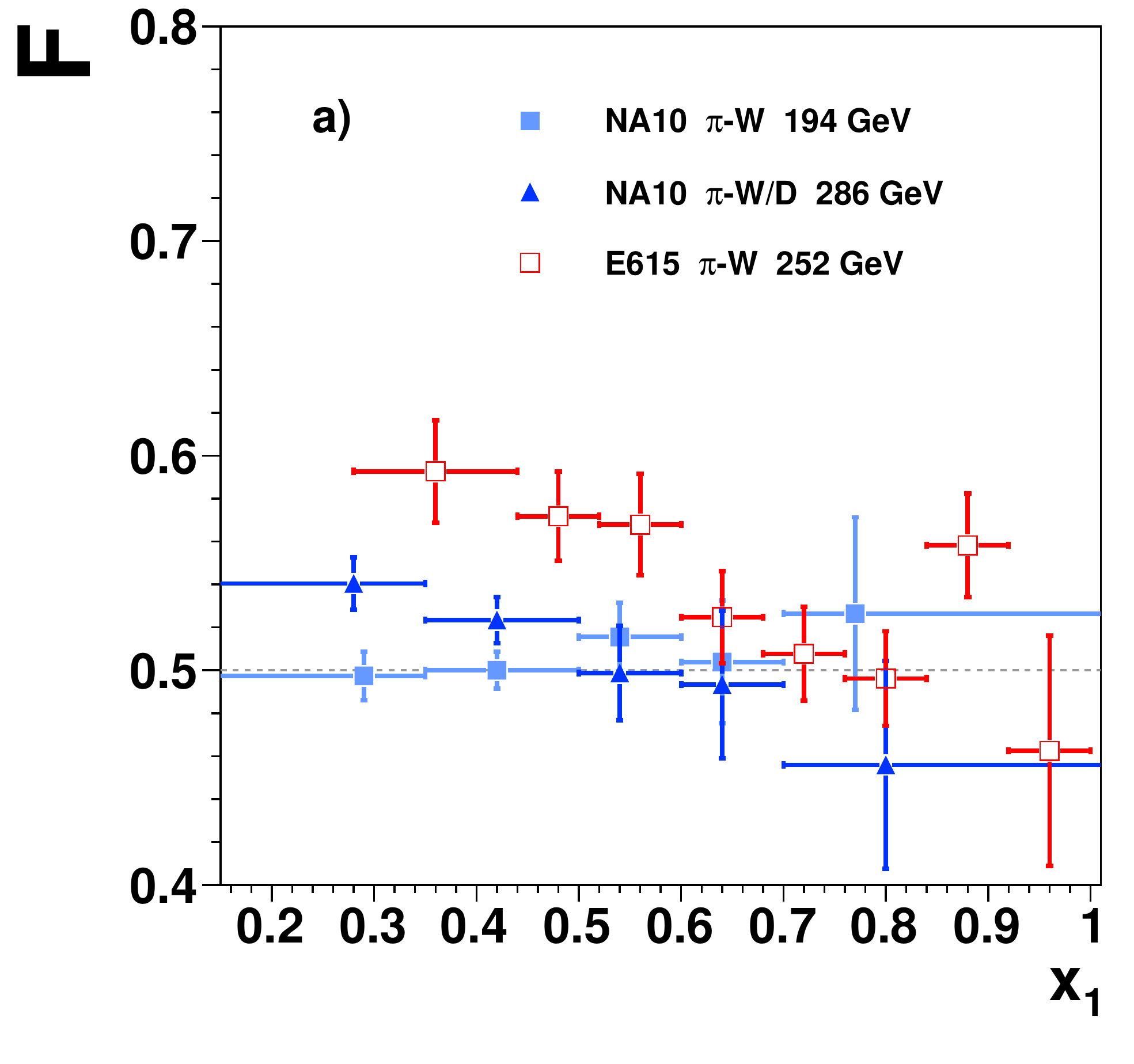}
\includegraphics[width=0.54\linewidth]{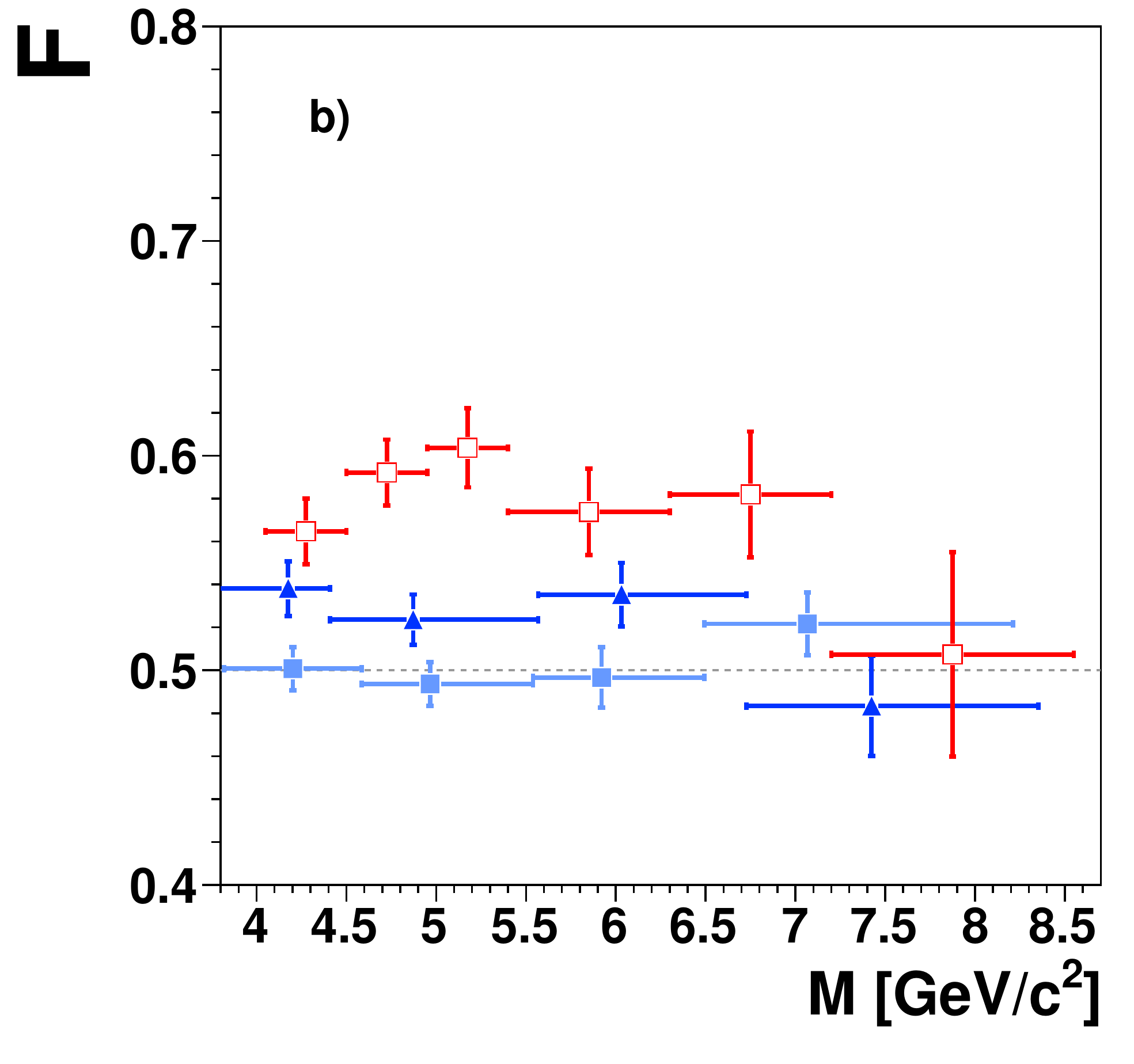}
\includegraphics[width=0.54\linewidth]{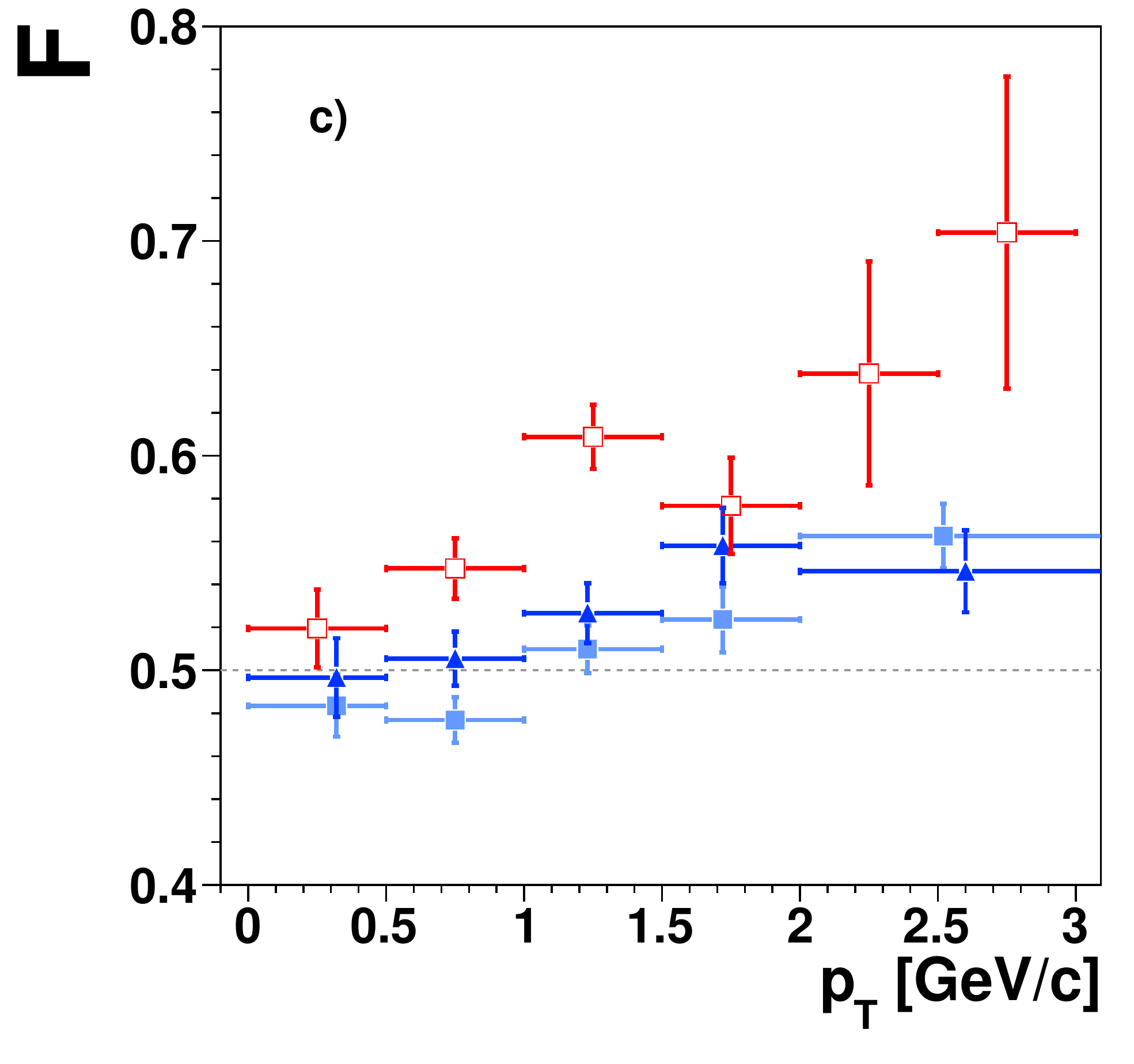}
\includegraphics[width=0.54\linewidth]{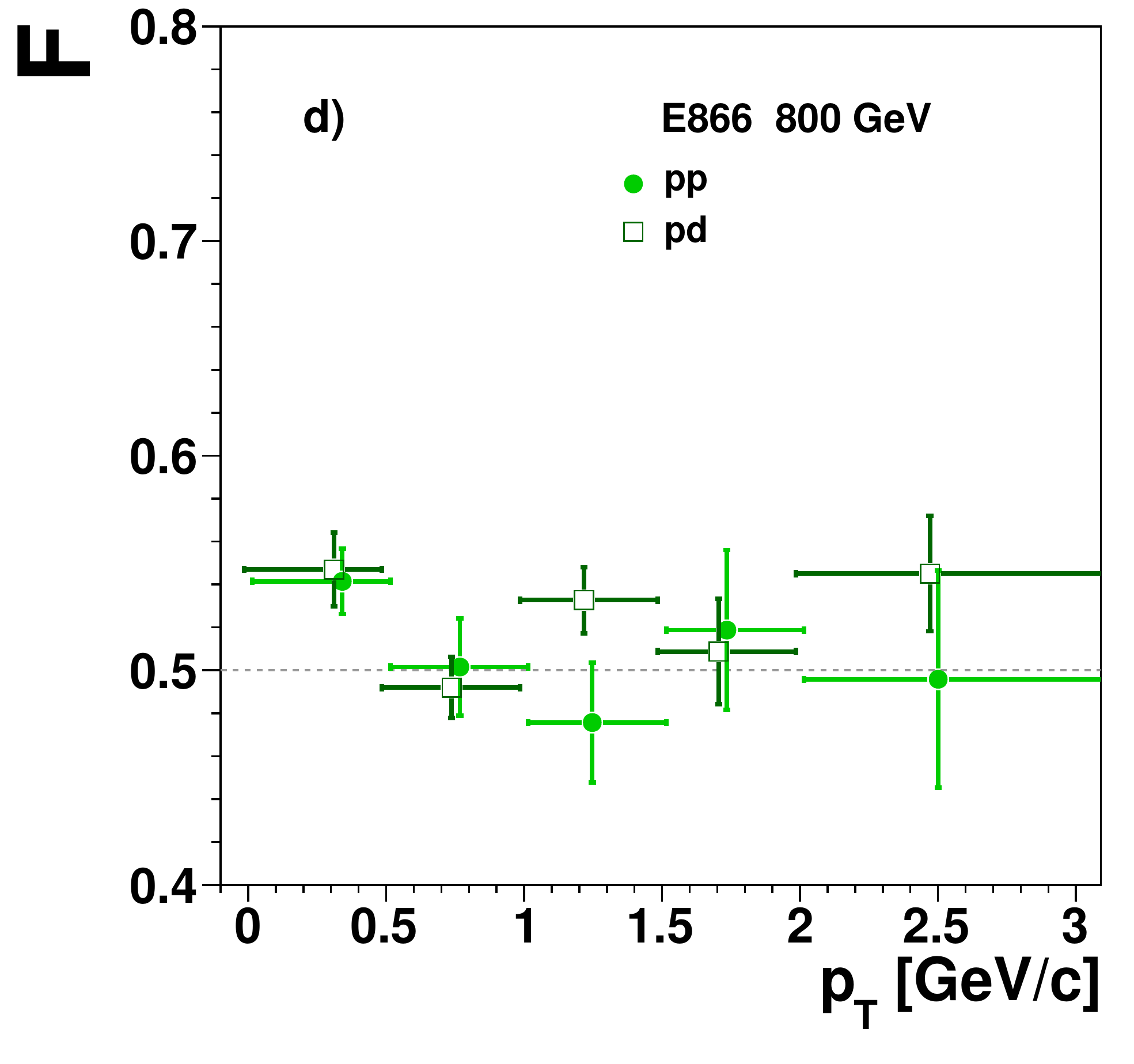}
\caption{The frame-invariant parameter $\mathcal{F}$ as a function of dilepton
kinematic variables, derived from Drell--Yan measurements obtained with pion
(a-c) and proton (d) beams. The E866 data points are slightly displaced in the
horizontal axis for improved visibility.} \label{fig:DYdata_F}
\end{figure}

Violations of the Lam--Tung relation in Drell--Yan angular distributions were
searched for in several experimental conditions. ``Anomalous'' effects were
evidenced in pion-nucleus experiments, which measured large azimuthal
anisotropies increasing with \pt\ and a strong reduction of the transverse
polarization at high $x_{1}$ (momentum fraction of the annihilating antiquark
in the beam pion). The largest effects, measured by E615~\cite{bib:E615}, are shown
in Fig.~\ref{fig:DYdata_lambdas}.
Figure~\ref{fig:DYdata_F} shows the values of $\mathcal{F}$ derived from the
angular distributions measured by E615, as well as by NA10~\cite{bib:NA10}, as
a function of $x_{1}$ (a), and of the dimuon mass $M$ (b) and transverse
momentum \pt~(c). The condition $\mathcal{F} = 1/2$ is increasingly violated
with increasing \pt, while there is no significant dependence on $x_1$ or $M$.
The panel (d) shows the E866 results, obtained in $pp$ and $pd$ interactions at
800~GeV~\cite{bib:E866}, perfectly consistent with the Lam--Tung expectation.
The most significant deviations from purely transverse dilepton polarization
are measured by E615 for $1 < p_{\rm{T}} < 1.5$~GeV$/c$ and $\langle x_{1}
\rangle \simeq 0.6$, in $\pi$-W collisions at 252~GeV, and by NA10 for $1.5 <
p_{\rm{T}} < 2$~GeV$/c$ and $\langle x_{1} \rangle \simeq 0.4$, in $\pi$-W and
$\pi d$ collisions at 286~GeV. The corresponding values of $\mathcal{F}-1/2$
are, respectively, $0.109 \pm 0.015$ and $0.058 \pm 0.018$.
%
%

Contrary to what the strong \pt\ dependence might suggest, we can easily
exclude that the enhancement of $\mathcal{F}$ with respect to $1/2$ is the
result of an event-by-event tilt between the polarization axis used in the
experimental analysis and the natural axis, caused by the intrinsic transverse
momenta of the partons. This observation can be easily understood by
considering the leading-order $1 + \cos^2 \vartheta_{\mathrm{CS}}$ distribution
and two extreme categories of events, in both of which the instantaneous
``natural'' axis (direction of the collision between partons) is significantly
tilted with respect to the Collins-Soper axis, but towards two orthogonal
directions: in one case the natural axis belongs to the plane of the colliding
hadrons, in the other case it belongs to the perpendicular plane. The first
type of events has a distribution characterized by $\lambda_\vartheta < 1$ and
$\lambda_\varphi > 0$, while $\mathcal{F}$, unaffected by rotations around the
axis perpendicular to the production plane, remains $1/2$. The second
event distribution is rotated with respect to the previous one by an angle
$\pi/2$ about the $z$ axis, implying that $\lambda_\varphi$,
the coefficient of the term $\sin^2 \vartheta \cos 2 \varphi$,
becomes negative, given that $\cos 2 (\varphi \pm \pi/2) = - \cos 2
\varphi$. On the other hand $\lambda_\vartheta$
remains unchanged (and smaller than $1$). Therefore, $\mathcal{F} < 1/2$. In
conclusion, parton transverse momenta lead on average to a reduction, rather
than an enhancement, of the overall observable anisotropy and, hence, of
$\mathcal{F}$.
The ``anomalous'' experimental observations cannot, therefore, be the
geometrical consequence of a deviation of the experimental axis from the
quantization axis of the elementary processes. They must reflect intrinsic
properties of the production mechanism, not properly described considering only
lowest-order perturbative processes.

Furthermore, we can also say that these anomalies cannot be caused by
relatively rare subprocesses, contributing as higher-order ``perturbations'' in
a standard QCD approach to the study of Drell--Yan production. We derive this
observation in the next lines, as a good example of the usefulness of the
formalism presented in this paper.
The maximum deviation of $\mathcal{F}$ from $1/2$ measured by E615,
$0.109 \pm 0.015$, allows us to deduce, using Eq.~\ref{eq:F},
that the fraction of dilepton events violating the condition $\mathcal{F}^{(i)}
\le 1/2$ is \emph{at least} as large as $0.22 \pm 0.03$.
Such a fraction would already have to be considered a very large contribution
to the Drell--Yan production yield, certainly not a ``perturbation''. However,
the real value must be even larger because this lower limit corresponds to an
extreme hypothesis: $\mathcal{F}^{(i)}$ is always either $1/2$, in the case of
standard $O(\alpha_s^0)$ and $O(\alpha_s^1)$ processes, or $1$, in the case of
the anomalous (hypothetical) ``higher-order processes'' violating the Lam--Tung
relation. In reality, the standard processes should have
$\mathcal{F}^{(i)}<1/2$, accounting for the parton transverse momenta effect
mentioned above, and the anomalous processes should have $\mathcal{F}^{(i)}$
values smaller than the extreme limit of $\mathcal{F}^{(i)}=1$. In these more
realistic conditions, in order to reproduce the E615 measurement the fraction
of Drell--Yan dileptons produced by the anomalous processes would need to be
comparable to the contribution of the ``lowest-order'' processes represented in
Fig.~\ref{fig:DYprocesses}.
Furthermore, such additional mechanisms, characterized by $\mathcal{F}^{(i)}$
values approaching $1$, would produce the dilepton in a rather uncommon angular
momentum state, where the $m=+1$ and $m=-1$ component amplitudes have
comparable magnitudes and interfere constructively, so as to maximize
$|a_{+1}^{(i)}+a_{-1}^{(i)}|$.
In particular, the limiting case $\mathcal{F}^{(i)} = 1$ corresponds to the
angular momentum state $\frac{1}{\sqrt{2}}
|\hspace{-.2em}+\hspace{-.2em}1\rangle + \frac{1}{\sqrt{2}} \,
|\hspace{-.2em}-\hspace{-.2em}1\rangle$, which is invariant by rotation around
the $y$ axis. As mentioned in Section~\ref{sec:constraints}, this specific
point of the phase space must be univocally attributed to the decay of a pure
eigenstate of $J_y$ with eigenvalue $0$ (top panel in Fig.~\ref{fig:triangles})
and it is impossible to reproduce it with a superposition of different states.
In conclusion, attributing the violation of the Lam--Tung relation in
pion-nucleus data to the existence of anomalous ``higher-order'' processes in a
perturbative QCD approach is equivalent to say that a \emph{very large}
fraction of Drell--Yan dileptons is produced in a fully (longitudinally)
polarized state \emph{with respect to the quantization axis perpendicular to
the production plane}.
This is an extremely peculiar and unrealistic scenario, requiring the existence
of a production mechanism that would lead, essentially in each and every
event, to a very exotic configuration of the dilepton spin.




\section{Summary}

The average angular momentum composition of a vector state is reflected in the
shape of its dilepton decay angular distribution
(Eqs.~\ref{eq:ang_distr_subproc}--\ref{eq:parameters}). The parameters of the
distribution can only take values inside a well-defined domain
(Eq.~\ref{eq:triangles}). For a specific mixture of production processes in a
given kinematic condition, there always exist a polarization observable
$\mathcal{F}$ (Eq.~\ref{eq:F}) independent of the choice of the quantization
axis (belonging to the production plane). The Lam--Tung relation represents the
particular case $\mathcal{F}=1/2$ (independent of production kinematics),
meaning that all subprocesses produce transversely polarized di-fermions with
respect to any polarization axis belonging to the plane of the colliding
hadrons. $\mathcal{F}$ can be determined from a single-variable distribution
(Eqs.~\ref{eq:cosalpha}--\ref{eq:lambdaalpha}), facilitating, in particular,
measurements of the violation of the Lam--Tung relation. The significant
violations of this relation found in pion-nucleus experiments cannot be
ascribed to the contribution of ``anomalous'' higher-order processes, because
rotational invariance and topological symmetry properties rule out a
``perturbative'' interpretation of the phenomenon.

\medskip

P.F., J.S.\ and H.K.W.\ acknowledge support from Funda\c{c}\~ao para a Ci\^encia
e a Tecnologia, Portugal, under contracts SFRH/BPD/42343/2007, CERN/FP/
109343/2009 and SFRH/BPD/42138/2007.


\end{document}